\documentclass[useAMS,usenatbib]{mn2e}
\usepackage{graphicx}
\usepackage{amsmath}

\newcommand{\ha}{H$\alpha$}

\newcommand{\neon}{[Ne~{\sc iii}]}
\newcommand{\argon}{[Ar~{\sc iv}]}
\newcommand{\helium}{He~{\sc i}}
\newcommand{\heliumb}{He~{\sc ii}}

\newcommand{\oi}{[O~{\sc i}]}
\newcommand{\niia}{[N~{\sc ii}]~6548\AA}
\newcommand{\niic}{[N~{\sc ii}]~5755\AA}
\newcommand{\oiiib}{[O~{\sc iii}]~5007~\AA}
\newcommand{\oiiia}{[O~{\sc iii}]~4959~\AA}

\newcommand{\nitrogen}{[N~{\sc ii}]}
\newcommand{\nitrogena}{[N~{\sc i}]}
\newcommand{\oxygen}{[O~{\sc iii}]}

\newcommand{\sulfur}{[S~{\sc iii}]}

\newcommand{\sulfurtt}{[S~{\sc i}]}

\title{``Distance mapping'' and the 3--D structure of BD +30 3639}

\author[S. Akras et al.]
{S. Akras$^{1}\thanks{e-mail:akras@astrosen.unam.mx}$ and
W. Steffen$^{1,2}$\\
$^{1}$Instituto de Astronom\'{\i}a, Universidad Nacional Aut\'onoma de
M\'exico, Ensenada, Baja California, M\'exico\\
$^{2}$Institut f\"ur Computergraphik, Technische Universit\"at Braunschweig,
Braunschweig, Germany}

\begin{document}

\date{Received **insert**; Accepted **insert**}

\pagerange{\pageref{firstpage}--\pageref{lastpage}}

\maketitle
\label{firstpage}

\begin{abstract}

BD +30 3639 is a member of a group of uncommon planetary nebula with
Wolf--Rayet central star and higher expansion velocities in \oxygen\ than
in \nitrogen\ lines. Images and high--resolution spectra from the literature
are used in order to construct a 3--D model of the nebula using the
morpho--kinematic code SHAPE. We find that two homologous expansion laws
are needed for the \nitrogen\ and \oxygen\ shell.
We conclude that the internal velocity field of BD +30 3639 decreases
with the distance from the central star at least between the \oxygen\ and
\nitrogen\ shells. A cylindrical velocity component is used to replicate
the high--speed bipolar collimated outflows. We also present a new kinematic
analysis technique called ``distance mapping''. It uses the observed proper
motion vectors and the 3--D velocity field to generate maps that can be used
as a constraint to the morpho--kinematic modeling with SHAPE as well as improve
the accuracy for distance determination. It is applied to BD+30~3639 using 178
internal proper motion vectors from Li et al. (2002) and our 3--D velocity
field to determine a distance of 1.52 $\pm$ 0.21 kpc.
Finally, we find evidence for an interaction between the
eastern part of nebula and the ambient $\rm{H_2}$ molecular gas.

\noindent
\end{abstract}


\section{Introduction}

Many surveys have been made in order to discover new Planetary Nebulae (PNe)
in our Galaxy (Parker \& Frew 2011 and references therein) showing a variety
of complex morphologies. The true 3--D dynamical structure of PNe has been
a problem of great complexity for many years. According to the theoretical
research, the deviation of PNe structure from the spherical symmetry occurs
for several reasons (i) the presence of strong magnetic fields and/or the
rotation of the central star (Garcia-Segura et al. 1997, 1999), (ii) the
effect of photo--dissociation (Garcia-Segura 2010) and (iii) the binarity of
central stars (Balick \& Frank 2002). Morphological and kinematical studies
of PNe help to understand their formation and evolution processes. High
resolution spatially resolved spectroscopy offers a useful tool to interpret
the structure as well as the kinematics of the nebula and to put limits to
dynamical models.

\begin{figure}
\includegraphics[scale=0.9]{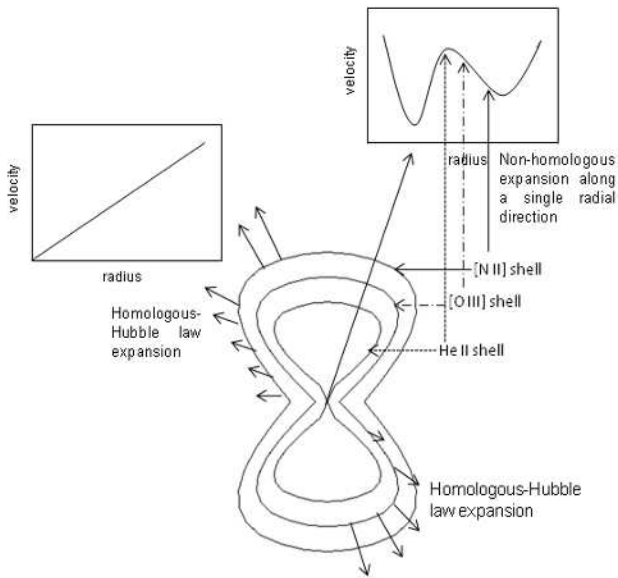}
\caption{A sketch illustrating the internal velocity field (``V--''or
``W--shape'') across the whole nebula (along a single radial direction) and
the homologous expansion law (Hubble--type) of a single shell (e.g. \nitrogen\
or \oxygen).}
\end{figure}

The issues of determining the 3--D structure and the internal velocity field
are strongly entangled. The projection on the sky flattens the structure
in images and the velocity field can not be uniquely determined from
Doppler-velocities and internal proper motion measurements unless restricting assumptions
are made. With some basic assumptions,
morpho-kinematical, radiation transfer modeling and hydrodynamical simulations
(or a combination of those) can provide a reasonable estimate of the 3--D structure
and velocity field. 

The internal velocity along a radial ray from the central star
through the nebula can show much more complex profiles (``V--'' or ``W--shape'')
than the commonly found velocity field of a single shell (Fig. 1). Such non-monotonic fields
have been reported in hydrodynamic models (Perinotto et al. 2004),
inferred with photoionization models (Gesicki et al. 2003, Gesicki \& Zijlstra 2003)
and observationally derived by Sabbadin et al. (2006). The expansion law
of the separate shells are usually more closely matched by the so-called "Hubble-law" or
homologous expansion. As investigated by Steffen et al. (2009), there may be
significant deviations from homologous expansion even within a single shell,
rendering the assumption of a homologous velocity field as insufficient for the
reconstruction of the 3--D structure. In such cases, additional information or
assumptions, such as some sort of global or local symmetry properties are required
for an unambiguous reconstruction.

Many PNe appear to show at least nearly a homologous expansion law based on
their optical images and PV diagrams (where it is safe to assume cylindrical symmetry).
High ionization ions with lower expansion velocities lay closer to the
central star than low ionization ions (Wilson 1950).
Nevertheless, there is a group of uncommon PNe which shows a
reverse behaviour (higher expansion velocities at high ionization ions
(e.g. \oxygen) than low ionization ions (e.g. \nitrogen)). This group includes BD+30~3639,
NGC 40, IC 351, NGC 6369, M1-32, NGC 1501, NGC 2022 and NGC 6751 (Medina et al.
2006). Most of them also display a non-spherical structure.

Regarding the morphology of PNe, the fraction of spherically and
non--spherically symmetric PNe in our Galaxy has been found at 19 per cent and
81 per cent, respectively (Parker et al. 2006; Miszalski et al. 2008). Spatially
resolved spectroscopic observations have revealed that the seemingly
spherically symmetric PNe can show a more complex structure along the line of
sight. An example is the Ring Nebula (Bryce et al. 1994, Guerrero et al. 1997,
Steffen et al. 2007) or BD +30 3639 (Bryce \& Mellema 1999). Therefore finding
the orientation of PNe poses a crucial problem to recognize their true 3--D
structure (Balick \& Frank 2002). Furthermore, the morphology of
PNe is likely to be related to the presence of magnetic fields or binary central
stars. In order to resolve the role of magnetic field and binarity in shaping
of PNe, it is necessary to study the complete 3--D structure and
velocity field in as many PNe as possible.

Knowledge of the 3--D structure and velocity field of PNe can also yield a
significant improvement in accuracy of distance determination, which still
poses a key problem in PNe research.
The lack of a reliable general method to determine the distance of PNe still
remains a great problem. Although, many attempts have been made to evaluate the
distance of PNe using several methods like the trigonometric parallax,
expansion parallax and statistical, their results are often discrepant
(Guzm\'an et al. 2009).

However, in the last few years, the expansion parallax method has become the most
accurate method for measuring the distance of particular PNe by using high
quality data from the interferometers or Hubble Space Telescope (HST) by
analyzing two epochs of images.

The accuracy of the expansion parallax method depends on (i) the non-spherically
symmetric expansion of PNe (Phillips 2005a) and (ii) the difference between
the angular expansion velocites on the sky and the material velocity (Mellema 
2004; Sch$\rm{\ddot{o}}$nberner, Jacob \& Steffen 2005). The authors claim that the measured
distance from expansion parallax method may be underestimated and has to be 
corrected by a factor of 1.1--1.5. Therefore, a detailed 3--D structure and 
velocity field are required to determine the distance with higher accuracy. 
With the 3--D morpho-kinematic code SHAPE (Steffen \& L\'{o}pez 2006, Steffen 
et al. 2011), in combination with suitable data, we are able to study 
simultaneously the rather complex 3--D structures and velocity fields of PNe.

The aim of this work is to obtain a comprehensive kinematic study of
BD+30 3639 (hereafter BD+30) using high-quality images and high--resolution
spectra from literature and to reconstruct a detailed 3--D morpho--kinematic
model. In section 2, a description of BD+30 is presented while in section 3,
we present the results from the morpho--kinematic modeling with SHAPE. The new kinematic
analysis technique "distance mapping" is presented in section 4. Finally, in
section 5, we discuss our results and we finish summing up our conclusions.

\section{BD +30 3639}

BD +30 is one of the most intensively observed PN in various wavelengths from
radio (Kawamura \& Masson 1996; Bryce et al. 1997), millimeter (Bachiller et
al. 1991, 2000), infrared (Latter et al. 1995; Matsumoto et al. 2008), optical
(Harrington et al. 1997, Bryce \& Mellema 1999; Li et al. 2002) to X-rays
(Arnaud, Borkowski \& Harrington 1996; Kastner et al. 2000; Maness et al. 2003;
Yu et al. 2009).

It is a young, compact and dense nebula with a bright Wolf-Rayet central star
(WC9) with effective temperature $\rm{T_{eff}}$ =42 kK and luminosity
$\rm{log(L/L_{\odot})=4.71}$ (Leunhagen, Hamann \& Jeffrey 1996). The mass-loss
rate and the wind velocity amount at $\dot{\rm{M}}$=1.3 $10^{-5}$ $M_{\odot}$
$y^{-1}$ and 700 km $\rm{s^{-1}}$, respectively, assuming a distance of 2.68 kpc
(Hajian, Terzian \& Bignell 1993). Its distance has been determined using
several methods such as the expansion parallax method and statistical methods
and the results vary between 0.67 kpc and 2.8 kpc (see Section \ref{distance_mapping}).

High--quality images from HST show a rectangular ring shape with low emission
at the central region. The emission is not uniform along its perimeter
revealing a brighter region in NE and a much fainter in SW direction (see Fig.
2). The dust in the neutral envelope forms the faint optical halo by
scattering the light from the central star (Harrington et al. 1997).
Furthermore, high-dispersion spectra from the William Herschel Telescope (WHT)
 at the level of detection show
an open--end shape in the \nitrogen\ lines and a closed shape in the \oxygen\
lines. The \nitrogen\ shell is slightly more
spatially and less spectrally extended than the \oxygen\ shell.
Bryce \& Mellema (1999), based on the same data, proposed that it is a prolate
nebula with low--density and fast moving polar regions, viewed almost pole--on
with the inclination angle at 20 degrees. Note that NGC 40, a planetary nebula
with WR--type central star as well, displays a similar structure. The dynamical
ages of NGC 40 and BD+30 are 4000 years and 600--800 years, respectively,
indicating the former as a more evolved version of the latter.

According to its expansion velocities, BD+30 belongs to a group of PNe which
shows a non--homologous expansion. In particular, the \oxygen\ expansion
velocities are higher compared to the \nitrogen\ ($35 \pm 1$ and $28 \pm 1$
km $\rm{s^{-1}}$, respectively (Bryce \& Mellema 1999); 39.5 and 21 km $\rm{s^{-1}}$, respectively,
(Medina et al. 2006)). The strong and extended X--ray emission from the hot
bubble of BD+30 ($T_X$=3 x $10^6$ K; Arnaud et a. 1996, Kastner et al. 2000),
indicates a strong wind--wind interactions, and/or a high pressure bubble that
drives the expansion. The intensity of X--ray emission increases from SW to NE.
This direction is consistent with that of CO bullets.

A pair of strong CO ($J$=2-1) bullets is oriented at position angles (PA) of 20 degrees
(blue--shifted) and 200 degrees (red--shifted), respectively, with projected expansion
velocities of 50 km $\rm{s^{-1}}$ (Bachiller et al. 2000). The presence of the CO bullets
supports the idea that the collimated outflows or strong collimated winds play
a role in the evolution and formation processes of BD+30. We note that
the kinematics of $\rm{H_2}$ molecular gas shows an opposite expansion behavior
to those of CO molecular gas, as it is expected for a prolate, almost pole--on
PN. Particularly, the NE part is redshifted by up to 60 km $\rm{s^{-1}}$ and the
SW part is blueshift by the same amount (Shupe et al. 1998).
Figure 2 presents images taken in different wavelength regions (optical,
$\rm{H_2}$ and CO).

\begin{figure}
\includegraphics[scale=0.45]{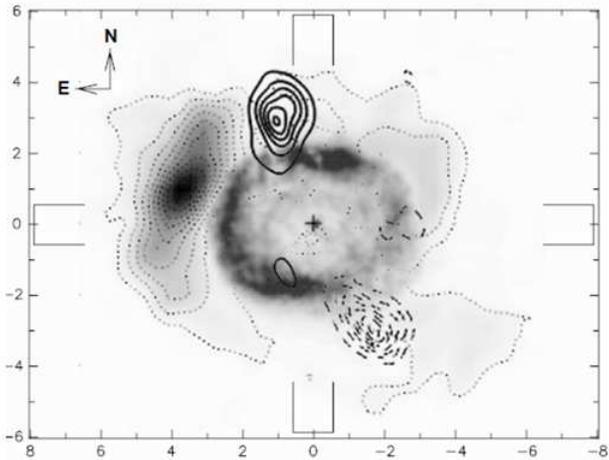}
\caption{Several types of maps have been combined: optical (Li et al. 2002 ), 
$\rm{H_2}$ (dotted contour, from Shupe et al. 1998) and CO (solid and dashed contours, 
from Bachiller et al. 2000).} 
\end{figure}

Regarding the internal velocity field across the nebula, Neiner et al. (2000)
tried to fit the nebular line profiles for various elements from different
stages of ionization, assuming a constant expansion velocity at 27 km
$\rm{s^{-1}}$ and a turbulent component at 15 km $\rm{s^{-1}}$. However, the
broad and split \oiiia\ line profile could not be reproduced properly since
the high-speed wings/shoulders are probably produced by the collimated outflows and not
by the expansion of the main nebula itself. Looking more carefully at the line
profiles of \neon, \sulfur, \nitrogena\ and \sulfurtt, one can discern
extended and faint wings similar to \oxygen\ line profile (see Fig. 4 in
Neiner et al. 2000). Hence, a simple constant velocity is not applicable
in BD+30 and additional structural and velocity components should be taken
into account in order to describe the outflows. These could be the molecular
bullets and their interaction with the main nebula.

Sabbadin et al. (2006) claimed that BD +30 is a rare object with
a "V-shape" velocity field. They found a decreasing function up to R=1.875 arcsec,
$\rm{V=110(\pm10)-43(\pm5) R}$ (R takes values between 1.725 arecsec and 1.875
arcsec) for the higher ionization ions (such as $\rm{O^{++}}$, $\rm{Ar^{++}}$,
$\rm{He^{++}}$ and $\rm{S^{++}}$) and an increasing velocity function
$\rm{V=15.0 (\pm 2) R}$ (R  takes values from 1.875 arcsec to 2.10 arcsec)
for the low ionization ions (such as $\rm{O^o}$, $\rm{O^+}$, $\rm{S^+}$
and $\rm{N^+}$) and producing a "V-shape" velocity profile.

The same "V--shape" velocity profile has already been found for 10 Galactic
PNe using the emission lines of \ha, \nitrogen\ and \oxygen\ (Gesicki et al.
2003). Moreover, Gesicki \& Zijlstra (2003) claim that they found
a more complex ''W--shape'' velocity field in three Galactic PNe
by fitting 9 nebular line profiles and covering the whole nebula from
high to low ionization ions. They also found that there is an anti--corellation
between the density distribution and the velocity field.
Responsible for that could be the stellar winds from the WR central star
of BD +30 which shows strong variations on a very short period of time (e.g. BD+30;
Grosdidier, Acker, Moffat 2000 \& R Hydrea; Zijlstra, Bedding \& Mattei 2002)
suggesting strong mass--loss fluctuations and a non-uniform density
distribution.

\begin{figure*}
\includegraphics[scale=0.48]{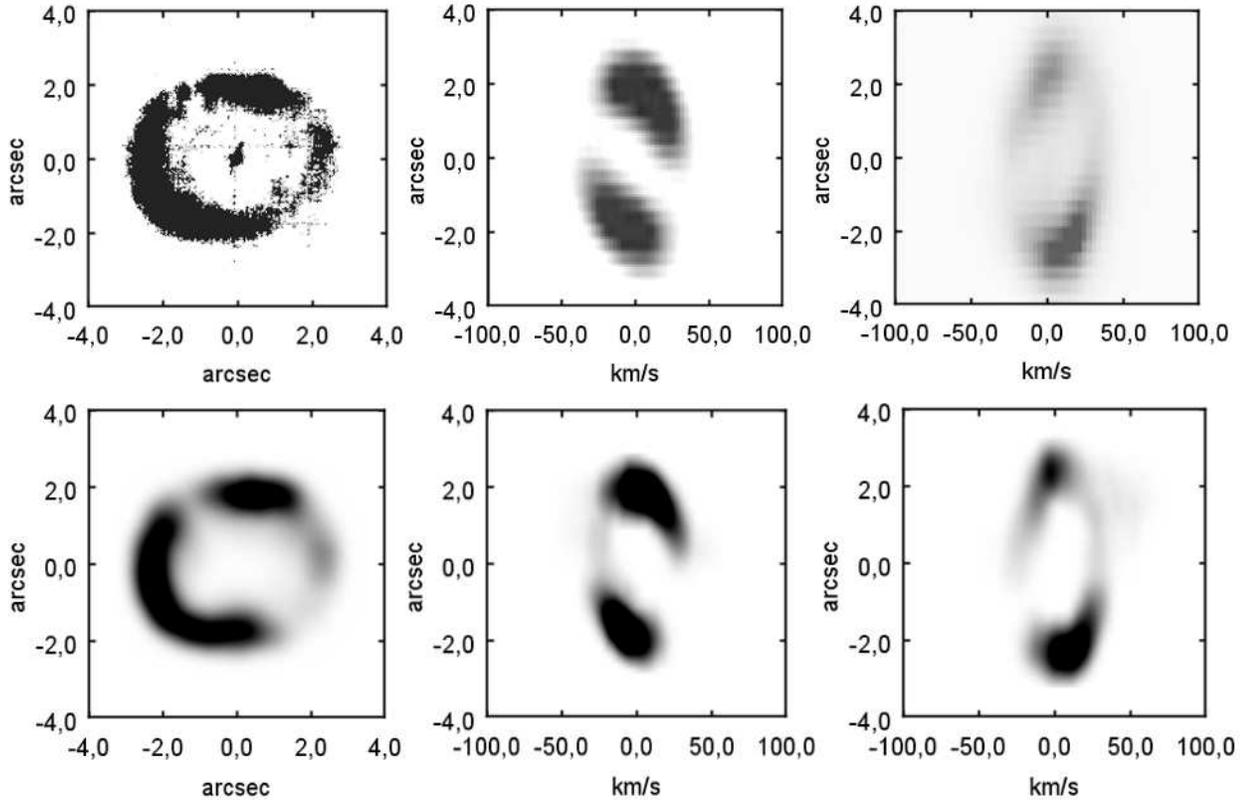}
\caption{The observed and modeled \nitrogen\ images on the left and the
corresponding synthetic PV diagrams of BD +30 3639 modeled with SHAPE.
The observations are at the top (Harrington et al. 1997, Bryce \& Mellema
1999) and the models at the bottom .
The middle panels show the PV line profile in north--south direction and the
right panels in west--east direction. The data are displayed with a square
root intensity scale to show at the same time the bright main nebula and
the faint high--velocity outflows. North is at the top and east at the left.}
\end{figure*}

\begin{figure*}
\includegraphics[scale=0.48]{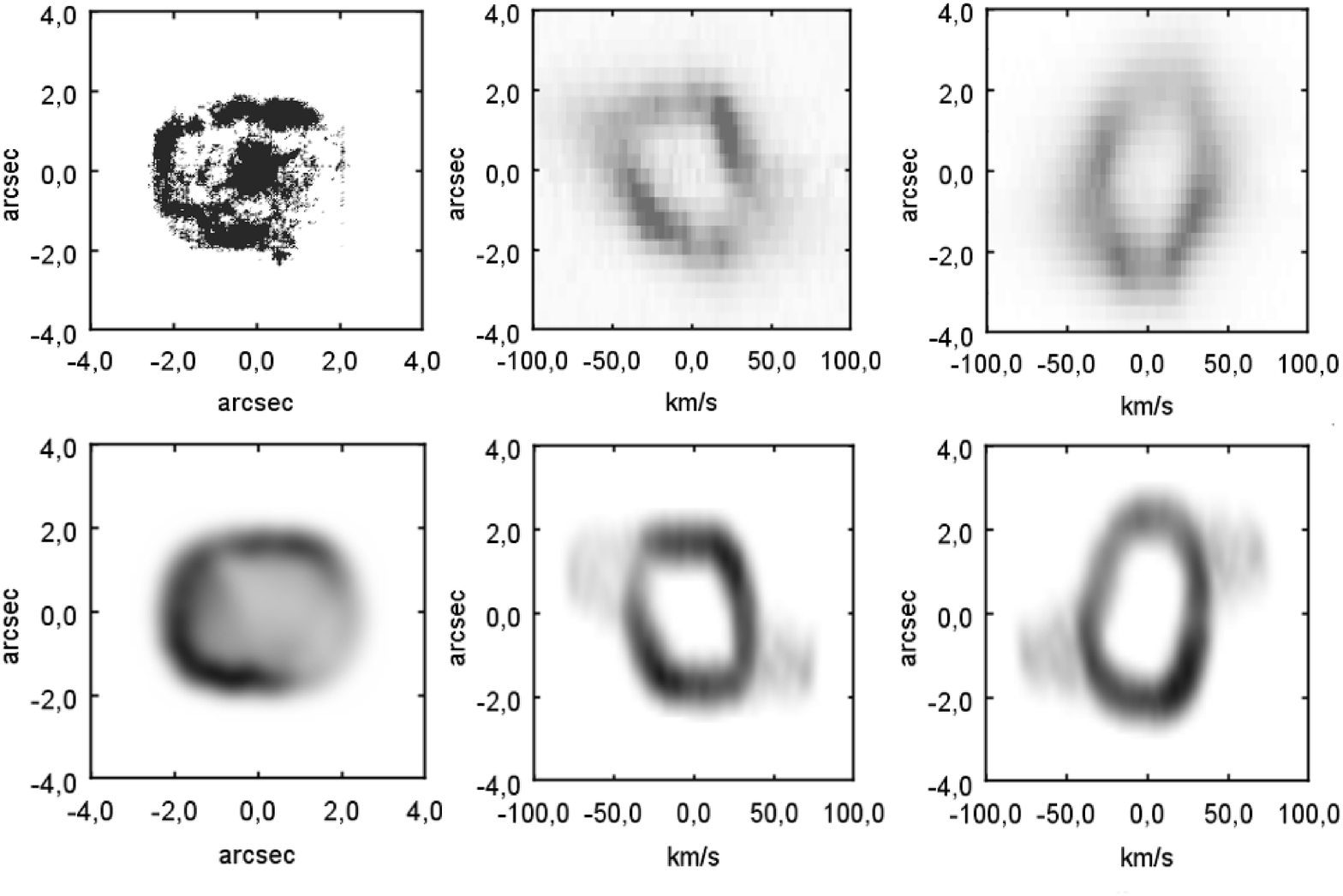}
\caption{The observed and modeled \oxygen\ images on the left and the
corresponding synthetic PV diagrams of BD +30 3639 modeled with SHAPE.
The observations are at the top (Harrington et al. 1997, Bryce \& Mellema 1999) 
and the models at the bottom
The middle panels show the PV line profile in north--south direction and the
right panels in west--east direction. The data are displayed with a square
root intensity scale to show at the same time the bright main nebula and
the faint high--velocity outflows. North is at the top and east at the left. }
\end{figure*}

\begin{figure*}
\includegraphics[scale=0.70]{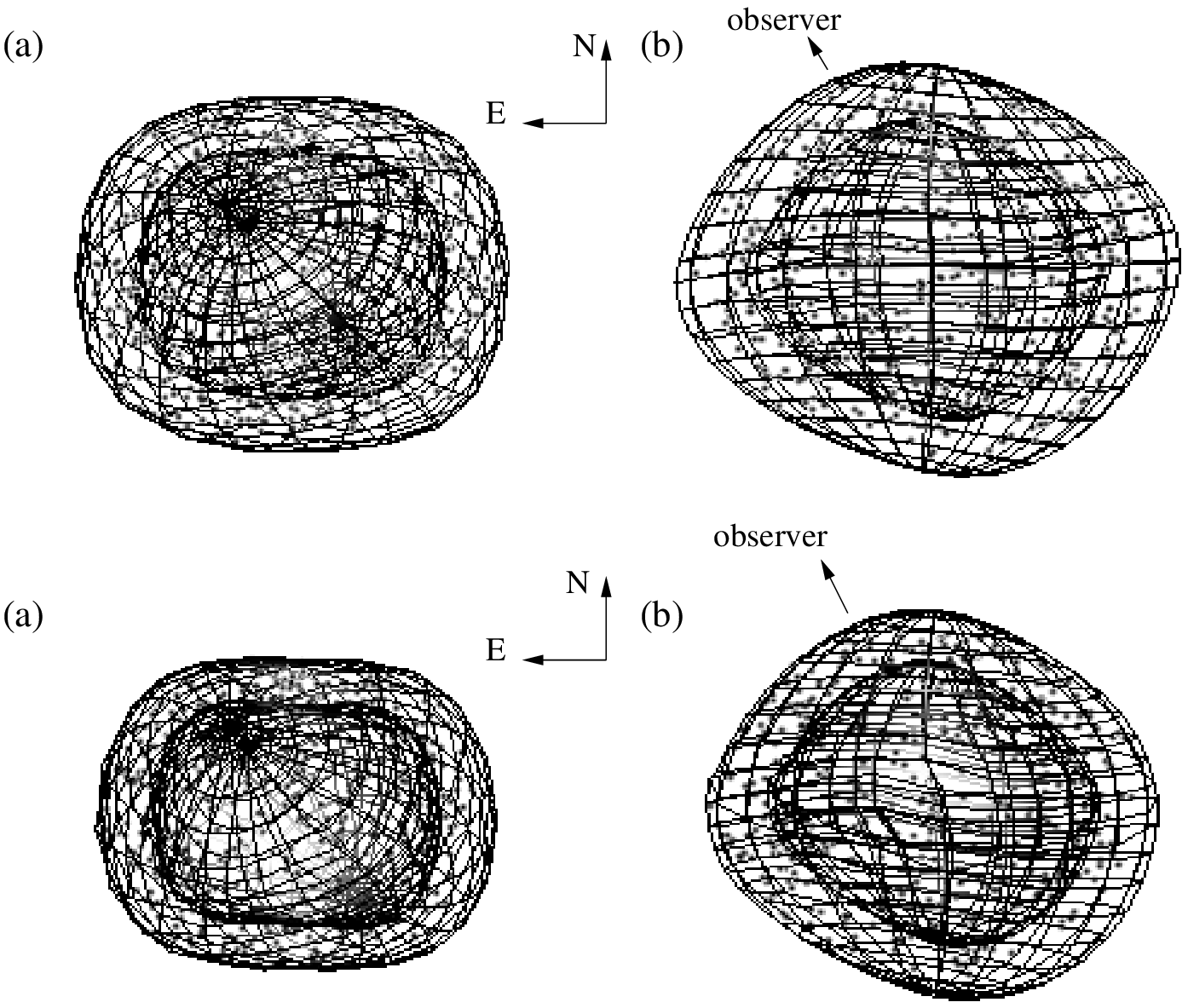}
\caption{SHAPE mesh model of BD +30 3639 for \nitrogen\ (upper panels) and
\oxygen\ (bottom panels) before rendering at two different oriantations.
(a) the view as seen from the observer (north up and east to the left) and 
(b) the view at 90 degrees inclination angle  (edge--on).}
\end{figure*}

\section{SHAPE model}

\subsection{Morpho--kinematic model}

In an attempt to improve previous distance determinations we construct a more
detailed morpho-kinematic model of the main nebula than those used before.
Earlier models consisted of an ellipsoidal shape with a homologous expansion
model (Li et al, 2002). We will include secondary deviations for this shape
and from the homologous expansion model. We introduce ``distance mapping''
as an additional constraint for the internal structure determination and to
improve the distance determination itself (see Section \ref{distance_mapping}).

In order to reconstruct the 3--D structure of BD +30, we use the morpho--kinematic
code SHAPE (Steffen \& L\'{o}pez 2006, Steffen et al. 2011). The observations are
optical images (Harrington et al. 1997), PV diagrams for \nitrogen\ and \oxygen\
lines from Bryce \& Mellema (1999) and internal proper motion measurements by
Li et al. (2002). The \oxygen\ shell reveals bright wings with expansion velocities
up to 80 km $\rm{s^{-1}}$ in both directions whereas the main nebula expands with
35--40 km $\rm{s^{-1}}$. In the PV diagrams, the \nitrogen\ shell is more extended
than the \oxygen\ shell which is consistent with the narrow-band optical images.
It expands with velocities 25-30 km $\rm{s^{-1}}$. The simultaneous reconstruction
of the \nitrogen\ and \oxygen\ PV diagrams was therefore not possible with one
homologous velocity law for both, with the \oxygen\ shell, nearly 10 per cent, closer
to the central star. Thus, two velocity laws are used (V=28 $\rm{r/r_o}$ km
$\rm{s^{-1}}$ for the \nitrogen\ line and V=40 $\rm{r/r_o}$ km $\rm{s^{-1}}$ for
the \oxygen\ line, where r is the distance from the central star and
$\rm{r_o}$ is the reference radius of the \nitrogen\ shell equal to 2.5 arcsec. 
Because the nebula is not spherically symmetric but shows a rectangular shape, 
the maximum distance ``r'' from the central star varies from 2.25 to 2.75 arcsec.
Therefore, the internal velocity field of BD +30 is a 
decreasing function of the distance from the central star over the range 
of the \oxygen\ and \nitrogen\ shell like in Figure 1.

The results of the rendered model are presented in Figures 3 \& 4 (lower panels) 
with the observed data (upper panels). Surprisingly similar PV diagrams have been 
found for NGC 6369 (Steffen \& L\'opez 2006) and NGC 6565 (Turatto et al. 2002). The 
latter shows a rectangular structure at almost the same inclination angle as BD +30 
and slightly more diffuse emission at the central region (see Fig. 16 in Turatto et 
al. 2002).

We begin with an ellipsoidal shell model similar to that used by Li et al. (2002).
Then we iteratively add detail, first to the structure and then to the velocity field.
This process is done manually as described in Steffen et al. (2011). The modeled
inclination and position angles are found to be 22$\pm$4 and 20$\pm$4 degrees, 
respectively, with the respect of the line of sight, which are consistent with the 
previous studies.

The upper panels of Figures 3 \& 4 show the HST narrow-band images (left panels; 
Harrington et al. 1997) and PV diagrams (middle and right panels; Bryce \& Mellema 1999) 
in \nitrogen\ and \oxygen\ emission lines, respectively. The main ionized shell shows an almost
rectangular shape with the major axis aligned from west to east and the minor axis
along the north-south direction. The emission is not uniform around the ring as a
brighter region appears in NE direction and a much fainter one in SW direction.

In the bottom panels of Figures 3 \& 4 we present the modeled images (left panels)
and PV diagrams (middle and right panels) in \nitrogen\ and \oxygen\ emission
lines from two different slit directions both going through the centre (NS and WE).
In particular, the \niia\ and \oiiia\ were obtained in NS direction (middle panels)
while the \niic\ and \oiiib\ in WE direction (right panels). The \nitrogen\ velocity
ellipse shows almost an open-ended shape in contrast to \oxygen\ velocity ellipse
which is closed. The open-ended shape of \nitrogen\ might just be a detection problem.
Figure 5 shows the mesh model of BD +30 for \nitrogen\ (upper panels) and \oxygen\
(lower panels) showing the geometry of the system before rendering at two 2 
different oriantations. Panels (a) display the view as seen from the observer 
and panels (b) display the view at 90 degrees inclination angle  (edge--on).

``Criss-cross'' mapping of the internal proper motion vectors shows extended and
elongated structure along the direction of the molecular outflows (see Section
\ref{criss-cross mapping}). This is indicative for a cylindrical velocity component
in the main nebula (Steffen \& Koning, 2011). Even if the AGB--wind is spherically
symmetric, an interaction with collimated outflows can produce an elongated elliptical
PNe (Espinosa et al. 2010) and impart such a cylindrical velocity component on the
main nebula. We therefore introduce a cylindrical velocity component that increases 
linearly away from the central star up to 30 km $\rm{s^{-1}}$ for the \nitrogen\ shell 
and 45 km $\rm{s^{-1}}$ for the \oxygen\ shell. This new velocity component 
is added to the already existing radial one in order to reproduce the high 
velocity bipolar outflows along the polar direction (Fig. 4, middle panels).
The cylindrical velocity component explains reasonably well the high velocity wings that 
Medina et al. (2006) referred and can not be reproduced by a simple homologous 
velocity law and a turbulent component.
Finally, there is also a small random velocity component of 4.0 km $\rm{s^{-1}}$ in each Cartesian
direction representing turbulent bulk motion and thermal broadening.

It is worth mentioning that the \nitrogen\ PV diagram reveals a velocity ellipse that is 
noticeably asymmetric in velocity and brightness which is hardly seen in \oxygen (Fig. 3, right panels). 
The western bright section shows a lower expansion velocities than the eastern counterpart. We
find that the model improves considerably when the velocity of the western region
is reduced by 20 km $\rm{s^{-1}}$. This asymmetry might have been caused by the
interaction of the western region of the nebula with the $\rm{H_2}$ molecular gas
around it. The $\rm{H_2}$ emission overlaps with the western part of the nebula,
suggesting that it might decelerate the expansion of the nebula in this direction
(see fig 1). In Fig. 6, we present the contour maps of modeled \nitrogen\ PV diagrams
with the velocity of the western region being 20 km $\rm{s^{-1}}$ less (6a; blue contours)
and without (6b; red contours) overlay of the observed PV diagrams (green contours).

\subsection{Criss-Cross mapping}
\label{criss-cross mapping}

\begin{figure}
\includegraphics[scale=0.75]{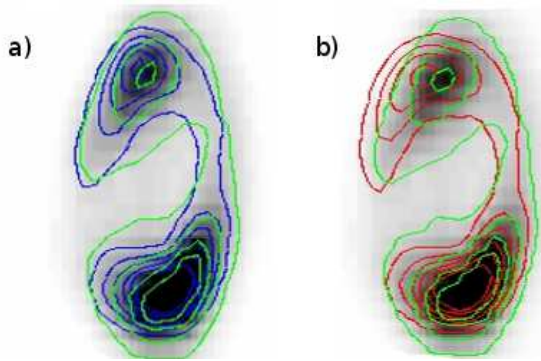}
\caption{The observed PV diagram of [N II] with the contours of modeled PV
diagram with the velocity of the western region being 20 km $\rm{s^{-1}}$ less
(a) and without (b). (See the electronic version for a color version of these
images.)}
\end{figure}

Steffen \& Koning (2011) introduced ``criss-cross'' mapping of internal proper
motion measurements and showed that it can help detect and interpret the 3--D
velocity field. Applying this new technique, they also showed that the velocity
field in BD+30 has deviations from a homologous velocity field. In this paper,
we reapply criss-cross mapping to 178 internal proper motion vectors published
by Li et al. (2002).

The observed criss--cross map of the \ha\ (Fig. 7, left panel) shows
clear indications that (i) the kinematic centre of the nebula is located
approximately 0.5 arcsec away from the central
star, even though the geometrical centre of the nebula is well centred on the star(s)
and (ii) there is a cylindrical velocity component along the NE-SW direction.
In contrast, the observed criss--cross map of the \nitrogen\ shows that the
kinematic centre of the nebula might be located closer to the central star
(Fig. 7, middle panel). The right panel in Figure 7, displays the modeled
criss--cross map including a 0.25 arcsec shift of the velocity field as 
deduced from the observed map.
Probably, the kinematic offset has been occured by a short event like a nova eruption from 
a binary system which has altered the homologous expansion by adding a new axisymetric component 
along the polar direction. RS Ophiuchi is a good example of a nova eruption with bipolar outflows 
(Sokoloski et al. 2008). The scenario of a nova eruption has alredy been proposed in order to 
explain the high Ne abundance of BD +30 (Maness et al. 2003) and Abell 58 (Lau et al. 2011). 
Accordingly, in this scenario, the nova eruption could change the velocity pattern of a spherically 
symmetric expanding PN, at least in some regions, converting the homologous expansion to a 
"V--" or "W--shape" along the radial direction from the central star(s).

\begin{figure*}
\includegraphics[scale=0.52]{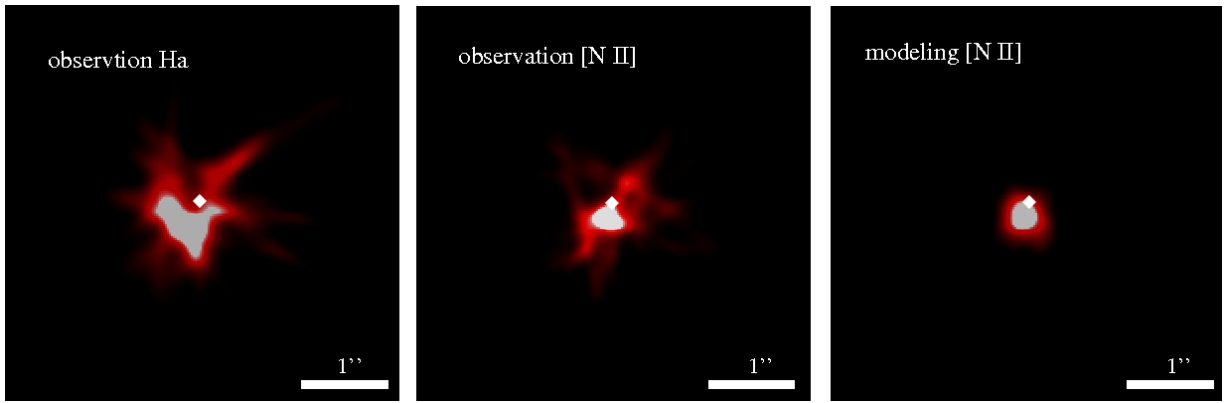}
\caption{The observed criss-cross maps of the H$\alpha$ and \nitrogen\
with the corresponding modeled criss-cross maps of the \nitrogen.
(See the electronic version for a color version of these figures.)}
\end{figure*}

From the Figures 3 \& 4 of Li et al. (2002), the angular deviation and the
position angles (PA) from the radial direction were determined by direct measurements
(Steffen \& Koning, 2011). In Figure 8, we present the angular deviation as a function
of PA for the \ha\ and \nitrogen\ lines.  Both do not follow a random distribution
around zero as would be expected for a radial expansion. Instead, they reveal a
sinusoidal pattern which can be explained by the offset of the kinematical centre.
However, the proper motion vectors of the \nitrogen\ line seem to lay closer to zero
suggesting a smaller offset, consistent with the criss-cross maps.
The difference between the \ha\ and \nitrogen\ proper motion vectors can scarcely
be discerned in Figure 8. For that reason, in Figure 9, we present histograms of
the observed and modeled angular deviation for PA between 155 and 335 degrees,
where the difference becomes noticeable. The upper panel shows the observed \ha\ and \nitrogen\
histograms and the lower panel shows the modeled \nitrogen\ histograms for
two offsets of the kinematic centre (0.50 arcsec; model M0.50 and 0.25 arcsec; model M0.25).
The observed \ha\ line is consistent with the model M0.50 (black histogram),
whereas the observed \nitrogen\ line is consistent with the model M0.25 (red histogram).
Particularly, the former show a peak between 15 and 20 degrees whereas the latter
show a peak between 5 and 10 degrees. The shift of the kinematic centre of the nebula
is confirmed for both lines. The value however, differs for \ha\ and \nitrogen\, probably
because they represent different regions in the nebula and the effect might be different
for them.

\begin{figure}
\includegraphics[scale=0.68]{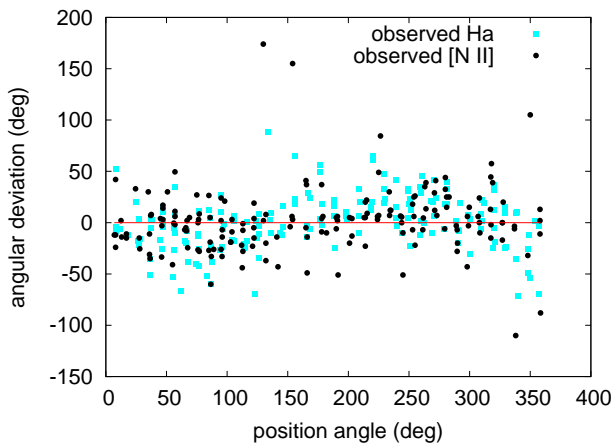}
\caption{The angular deviation from the radial direction of the internal proper
motion vectors is plotted against the position angle. The observed \ha\ data are in blue
and the \nitrogen\ are in black (Li et al. 2002). (See the electronic version for a color 
version of this figure.)}
\end{figure}

\begin{figure}
\includegraphics[scale=0.65]{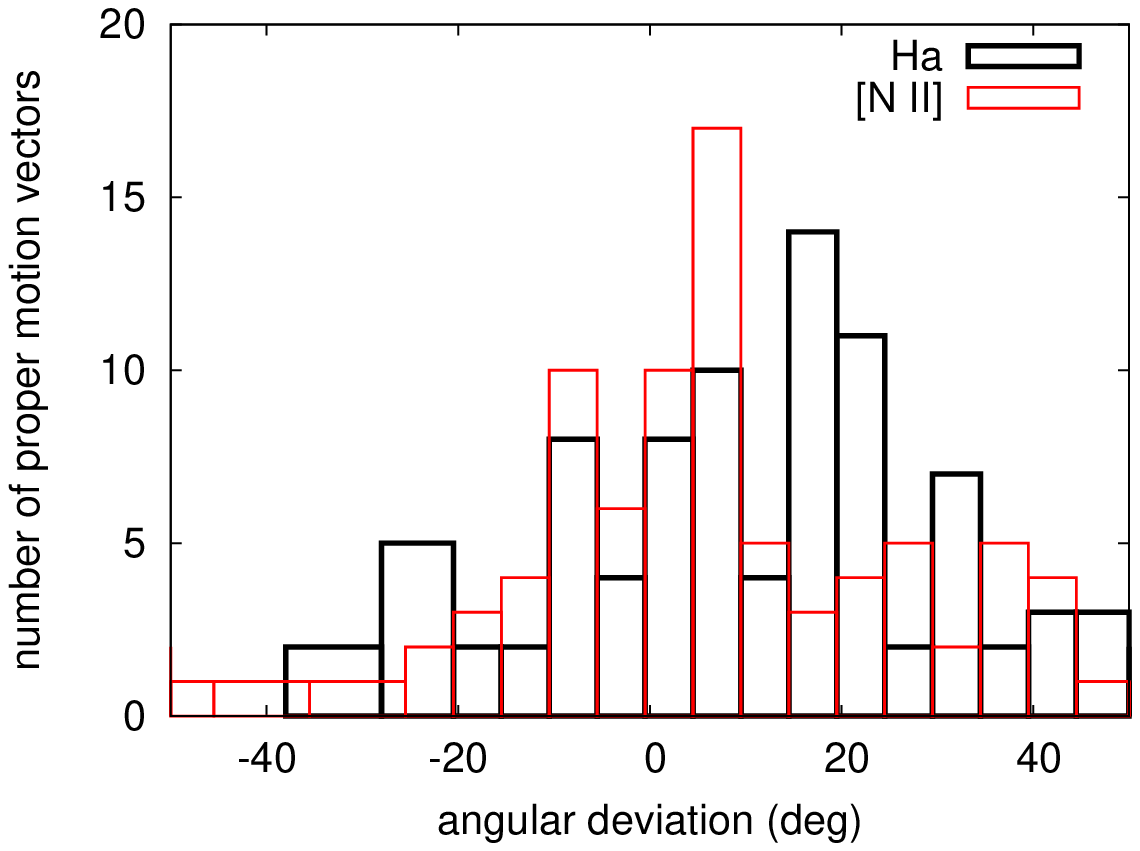}
\includegraphics[scale=0.65]{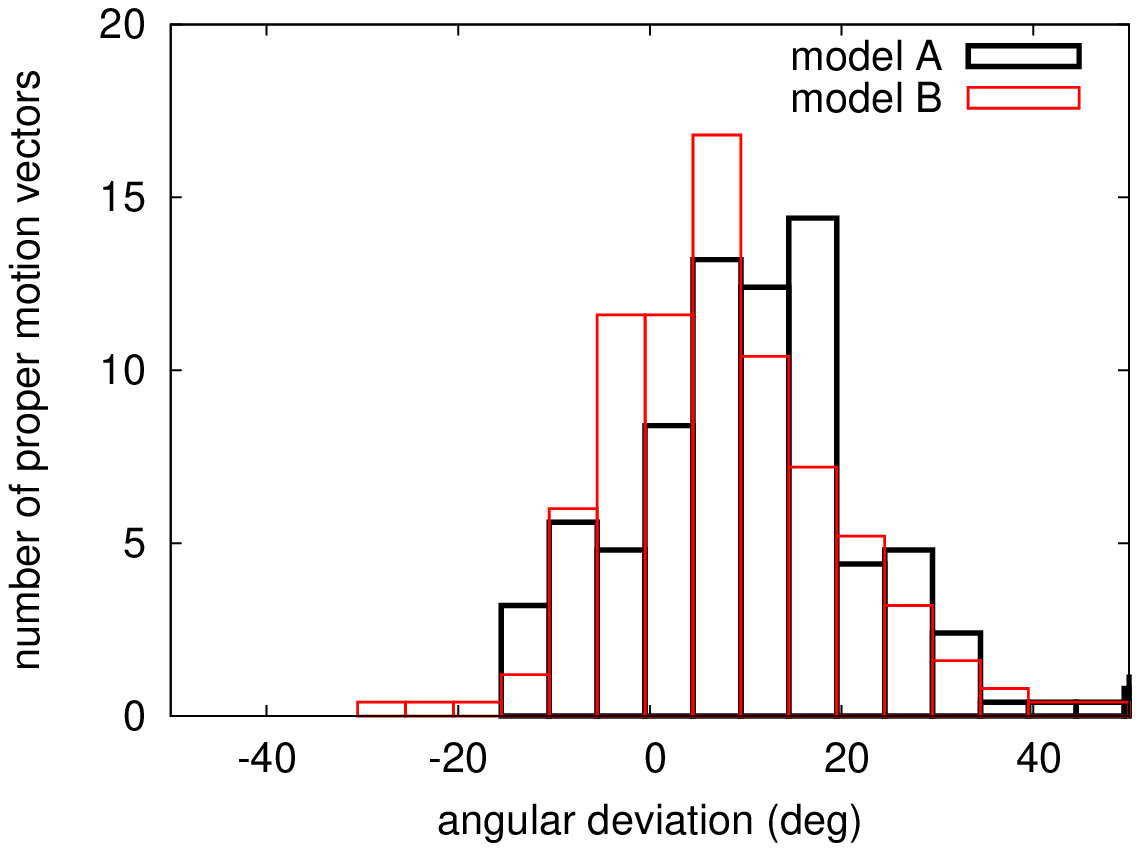}
\caption{The angular deviation histograms of the observed \ha\ (black color) and
\nitrogen\ (red color) data (upper panel) and of the modeled \nitrogen\ data for
two offsets of the kinematic centre of the nebula, 0.50 arcsec (black color; model M0.50)
and 0.25 arcsec (red color; model M0.25) (lower panel) for PA between 155 and 335.
(See the electronic version for a color version of these figures.)}
\end{figure}

\section{Distance mapping technique}
\label{distance_mapping}

The distance parameter is a key problem in PNe research. Knowledge of the
observed angular expansion rate and the true internal 3--D velocity field should
yield a more precise and direct determination of the distance. This information
can be derived from observations of internal proper motion using high quality
data either from interferometers or HST and analyzing two epochs of images and
Doppler-shift of the gas, combined with a 3--D morpho-kinematic model.

\begin{figure*}
\includegraphics[scale=0.50]{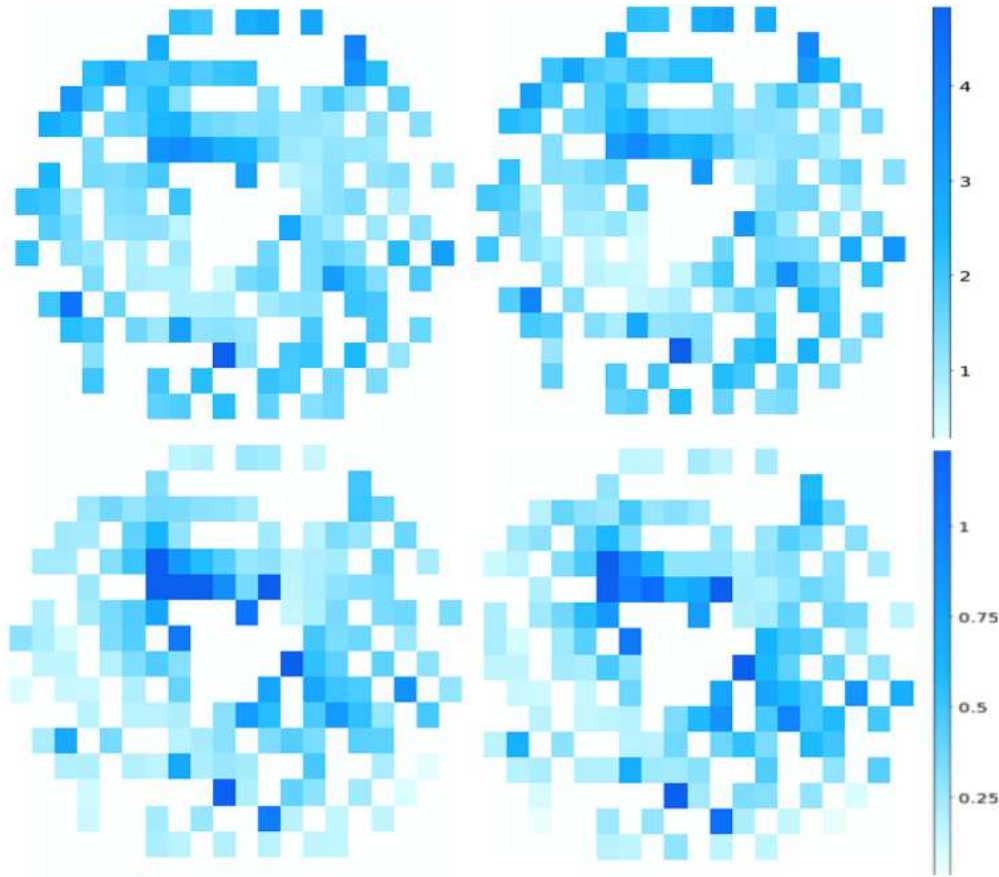}
\caption{The distance maps (upper panels) and the error maps (lower panels) of
BD +30 for two models with different offsets of the kinematic centre of the nebula 
(left panels: model M0.50 and right panels: model M0.25). The scale bars orrespond to 
the ranges indicated in units of kpc. (See the electronic version for a color version of these maps.)}
\end{figure*}

The basic idea is to determine the distance for small boxes throughout
the 2--D image of a nebula using as many proper motion vectors as possible and
generate a map of the distances obtained. In the ideal case, obviously, 
the distance for each box should be
the same and the resulting distance maps should result in a uniform
value within the noise limits. Any systematic deviation hint
towards deviations of the model structure and/or velocity field from the actual fields.
The distance map can therefore be used as a constraint for the
morpho--kinematic modeling as well as in determining distance with better
accuracy.

In order to determine the distance of BD +30, we use our 3--D model velocity
field based on the PV diagrams from Bryce \& Mellema (1999) and 178 internal
proper motion measurements from Li et al. (2002). We divide the
nebula's image in small boxes and calculate the mean value of proper motion
and tangential velocities of each one. Then, the distance and the standard
deviation are calculated for each box, by means of equation (1),

\begin{equation}
\rm{d} = 211 \frac{\rm{V_t}} {\dot{\rm{\theta}}},
\end{equation}

where d is the distance in kpc, $\rm{V_t}$ is the modeled tangential velocity in
km $\rm{s^{-1}}$ and $\dot{\rm{\theta}}$ is the observed proper motion in
mas/yrs. From these data we obtain a distance for each box and then construct
map of the  distance and its formal error (Fig. 10). Finally, we calculate the weighted 
mean value to find the best distance and the corresponding statistical error $\sigma$.

Figure 10 presents the distance maps for two models with different offset of
the kinematic centre of the nebula (left upper panel: model M0.50 and right upper panel: 
model M0.25) and the corresponding error maps for each model (lower panels).

According to the distance maps, there are two region, where the resulting distances are
systematically less (region A) and higher (region B) compared with
the rest of the nebula (Fig. 11). In particular, the distances in the region A
are determined 20--30 per cent less by the model M0.50 (mean value = 0.67 kpc) compared with 
the model M0.25 (mean value = 0.92 kpc). 
In the region B, the distances do not change significantly for different kinematic centre ($<$ 10 per cent).
Figure 12 shows the observed angular expansion velocities of \ha\ and 
\nitrogen\ lines as a function of position angle. One can see that, although both lines 
cover the same angular expansion velocities, there is a region between PA=100 and 135 
degrees (region A), where the \nitrogen\ line shows systematically higher angular expansion 
velocities than the \ha\ line, suggesting a possible systematic error. The overestimation of 
the angular expansion velocities in that region results in underestimating of the distance 
determination. It is therefore, not clear whether they represent true features in the nebula, 
are deficiencies in the model or systematic error in the proper motion measurements. 
Probably, a more complex velocity field might be required with an additional 
velocity component along the direction of the bullets. We investigate this possibility below.
Future observations might be needed to verify the complexity of the velocity field.

The resulting distances in the region B are higher than the rest of the nebula, due to
the low angular velocities (see Fig. 12;  $-30 < PA < 50$).
We speculate that (i) the velocity field of BD +30 is even more complicated with
significant deviations from homologous expansion within each shell,
or (ii) the current kinematic data might have systematic errors and do not help 
to constrain a more complex model.

The option of a more complex velocity field is investigated by introducing the effect 
of an additional cylindrical velocity component in a limited region around the direction
of the molecular outflows. Higher cylindrical velocities give higher
distances, but still within the errors. The cylindrical component does not seem
to influence significantly the final distance measurement. We, finally, evaluate the distance
of BD +30 at 1.58$\pm$0.21 kpc and 1.46$\pm$0.21 kpc from the model M0.50 and M0.25, respectively.
However, from previous distance determinations, we point out that the
distance of BD +30 is clearly less than 3.0 kpc and therefore values
higher than 3.00 kpc (12 points or 8 per cent) could be excluded from our calculation. In this case,
the distance is found to be 1.52$\pm$0.21 kpc and 1.40$\pm$0.20 kpc for model M0.50 and M0.25, respectively. 
In addition, if regions A and B are excluded from the distance determinations, we find it   
to be 1.55$\pm$0.21 kpc and 1.43$\pm$0.20 kpc for the models M0.50 and M0.25, respectively. 
All the resulting distances are within the errors and consistent with the previous studies.

\begin{figure}
\includegraphics[scale=0.45]{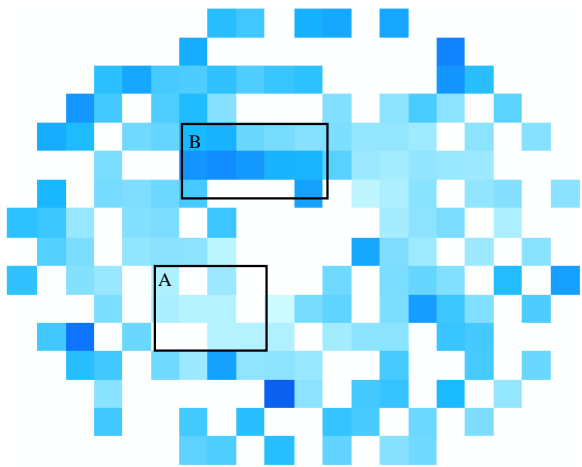}
\caption{The distance map of the model A with the regions A and B.
(See the electronic version for a color version of this map.)}
\end{figure}

\begin{figure}
\includegraphics[scale=0.68]{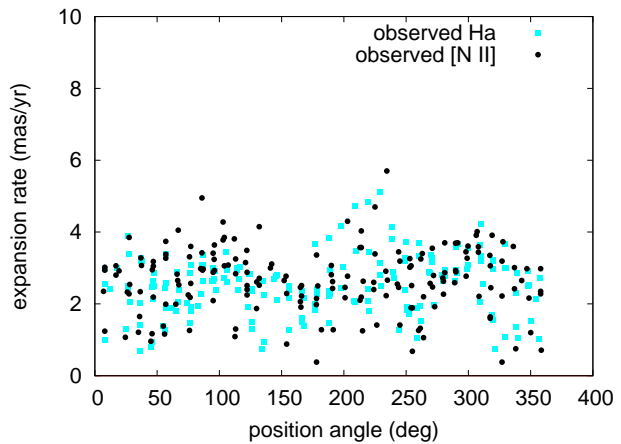}
\caption{The propor motion rate distribution is plotted against the position angle.
The observed \ha\ data are in blue and the \nitrogen\ data are in black
(Li et al. 2002). (See the electronic version for a color version of this figure.)}
\end{figure}

Previous distance determinations of BD +30 have been performed by using several methods,
such as the parallax methods in radio and optical wavelengths and statistical methods.
In Table 1, we give all the measured distances of BD +30. The results from the statistical
methods vary significantly by a factor of 2 or 3, whereas the recent distance measurements 
from the parallax method give more accurate results. Our resulting distance is determined by
using almost 178 proper motion vectors and a complex 3--D velocity field improving
the accuracy of the distance.

\begin{table}
\centering
\caption[]{BD +30 distances derived by different methods.}
\label{table4a}
\begin{tabular}{llllllllllll}
\hline
{\em \rm{Distance}} & optical/radio & reference\\
(kpc)   & parallax/statis. &\\

\hline
2.80$\pm$1.37 &  r/p &  Masson 1989\\
2.68$\pm$0.81 &  r/p &  Hajian et al. 1993\\
1.50$\pm$0.40 &  r/p &  Kawamura \& Masson 1996\\
1.20$\pm$0.12 &  o/p &  Li et al. 2002\\
2.57          &  s   &  Phillips 2005\\
0.67          &  s   &  Phillips 2002\\
1.85          &  s   &  Zhang 1995\\
2.14          &  s   &  Phillips 2004\\
1.16          &  s   &  Cahn et al. 1992\\
0.73          &  s   &  Daub 1982\\
1.84          &  s   &  Van de Steene \& Zijlstra 1994\\
1.46$\pm$0.21 &  o/p+model &  present work\\
\hline
\end{tabular}
\medskip{}
\end{table}

\section{Discussion and Conclusion}

Our 3--D morpho--kinematic model showed that the simultaneous reconstruction
of \nitrogen\ and \oxygen\ PV diagrams is not possible with a single
homologous velocity field across the nebula. BD +30 is an uncommon PN
with higher \oxygen\ expansion velocities than \nitrogen\ (Bryce \& Mellema
1999; Sabbadin et al. 2006; Medina 2006).

We found that two velocity laws are necessary to model the whole
structure of BD +30 (V=28 $\rm{r/r_o}$ km $\rm{s^{-1}}$ for the \nitrogen\ line
and V=40 $\rm{r/r_o}$ km $\rm{s^{-1}}$ for the \oxygen\ line).
We concluded therefore that the internal velocity field of BD +30 is a
decreasing function of the distance from the central star over the range
of the \oxygen\ and \nitrogen\ shell. This is consistent with some of the hydrodynamical
models of PNe (Perinotto et al. 2004) that show complex velocity profiles with
``V--''or ``W--shapes''. Therefore, supplementary spectroscopic observations from
\helium, \heliumb, \nitrogena, \oi\ and \argon\ are required in order to cover the
inner and the outer parts of BD +30 to obtain a more complete 3--D
morpho--kinematic model.

We obtained a 3--D morpho--kinematic model that includes the main
morphological and kinematical characteristics reproducing key features 
of the 2--D images and the position--velocity diagrams. The fast--moving collimated 
bipolar outflows were replicated by a cylindrical velocity component in polar 
direction. The inclination and position angle were found to be 22$\pm$4 and 
20$\pm$4 degrees respectively, with the respect to the line of sight. 
We also found that the \nitrogen\ velocity ellipse in the west--east slit direction 
is non--symmetric in velocity and displays low expansion at the western region, 
probably due to the interaction of main nebula with the ambient $\rm{H_2}$ molecular cloud.

A new kinematic analysis technique called ``distance mapping'' was developed
based on the 3--D velocity field and internal proper motion measurements.
The distance of BD +30 was found to be 1.46$\pm$0.21 kpc in the case of a kinematic 
offset of 0.25 arcsec. Since, our kinematic model was built using 
high resolution images, PV diagrams for two slit position and 178 proper
motion vectors, we expect that our resulting distance is more accurate
than the previous studies. 
In addition, the distance mapping technique can be used to constrain    
morpho--kinematic models. Some problems with fitting a suitable model to the proper  
motion data hint towards a possible systematic error in the proper motion data. 
Remeasurement of the proper motion data based on the existing or new ones might be required 
to resolve any systematic errors. As the time baselines for high spatial
resolution observations increases, detailed internal proper motion measurements 
can be obtained for many objects. Distance mapping can then be expected contribute 
significantly to the improvement of distance determination and to constrain 
3--D modeling of the objects.
\\
\\
\\
\\
\\
This work has been supported by grant from UNAM PAPIIT IN 100410. S. A.
acknowledges a postdoctoral scholarship from UNAM-DGAPA. We also would 
like to thank the referee, M. Lloyd, for her valuable comments and 
suggestions. 

\bibliographystyle{mnras}

\clearpage

\end{document}